\documentclass[12pt]{article}
\usepackage{amssymb}
\usepackage{color,graphicx}
\usepackage[dvipsnames]{xcolor}
\usepackage{amsmath}
\usepackage{epsfig} 
\usepackage{enumerate}
\usepackage{bm}
\def\eps{{\varepsilon}}

\def\S{\mathcal{S}}

\def\eRM{{\mathrm e}}
\def\dRM{{\mathrm d}}

\def\my{{\bm y}}

\def\mx{{\bm x}}
\def\mv{{\bm v}}

\def\mk{{\bm k}}

\def\mpp{{\bm p}}

\def\mq{{\bm q}}

\def\eps{\varepsilon}
\def\boldnabla{{\bm \nabla}}

\newcommand{\psid}{\widetilde{\psi}} 

\allowdisplaybreaks
\allowbreak

\begin{document}

\title{ Passive advection of percolation process: Two-loop approximation}

\author{\v{S}arlota Birn\v{s}teinov\'{a}$^{1}$, Michal Hnati\v{c}$^{1,2,3}$,
  Tom\'{a}\v{s} Lu\v{c}ivjansk\'{y}$^{1}$, \\ Luk\'{a}\v{s}~Mi\v{z}i\v{s}in$^{2,3}$,
    Viktor \v{S}kult\'ety$^{4}$}

\maketitle              
\mbox{ }\\
$^1$ Pavol Jozef \v{S}af\'{a}rik University, \v{S}rob\'{a}rova 2, 
041 54 Ko\v{s}ice, Slovakia \\
$^2$ Joint Institute for Nuclear Research, 141980 Dubna, Russia \\
$^3$ Institute of Experimental Physics, Slovak Academy of Sciences, Watsonova 47, 
040 01 Ko\v{s}ice, Slovakia\\
$^4$ SUPA, The University of Edinburgh,
Peter Guthrie Tait Road, Edinburgh EH9 3FD, United Kingdom

\begin{abstract}
The paradigmatic model of the directed percolation process is studied near its second order
phase transition between an absorbing and an active state. The model is first expressed in a form
of Langevin equation and later rewritten into a field-theoretic formulation. The ensuing response
functional
is then analyzed employing Feynman diagrammatic technique and perturbative
 renormalization group method. Percolation process is assumed to occur in external
 velocity field, which has an additional effect on spreading properties. Kraichnan rapid change ensemble
 is used for generation of velocity fluctuations. The structure of the fixed points structure is obtained
 within the two-loop approximation.
\end{abstract}

\section{Introduction}
In almost every realm of everyday life
 physical systems under non-equilibrium conditions are encountered. 
Mutual interplay between dissipative and driving forces  
give rise to a complicated and intriguing macroscopic behavior
\cite{Zia95,HHL08,Tauber2014,Krapivsky}. Among most interesting, 
 and at the same time very difficult to be tackled theoretically, are 
 systems far from thermal equilibrium. 
Despite a lot of effort that has been made during last decades, fundamental understanding
  of non-equilibrium physics is still missing.
  
Reaction-diffusion problems appear commonly in biological systems and due to its
very nature they could not be described by equilibrium statistical physics. 
 Spreading constituents (atoms or more generally agents) interact with each other and
 thus the number of agents is not conserved. As control parameters are changing, it might happen
that a underlying reaction scheme allows an existence of so-called absorbing state. Once system 
enters this state,
 it could not leave it. Clearly, this causes the system to be non-ergodic and thus impossible to study by
 equilibrium methods. 
  It might happen that absorbing state is separated from an active state, i.e.,
    a state with fluctuating (non-zero) constituents, by a critical point. At this point an
    emergent scaling behavior is observed, quite analogous to critical points in equilibrium 
    systems \cite{Zinn,Amit}.
It is well-known that at critical point
 large scale spatio-temporal fluctuations govern the overall statistical properties and
 the resulting collective behavior can be effectively described by a certain set of continuous fields. The usual
manifestation of criticality as a presence of divergences in various correlation functions is expected.
 A classical example is provided by the directed percolation process (DP), also known
as Gribov process in hadron physics \cite{HHL08,Stauffer,Cardy80}. DP is mainly used as
a simple model for a description of a population dynamics on the edge of extinction. 
Other possible applications embody high-energy physics, 
 fluid turbulence, ecology and others \cite{Tauber2014,Odor04,Janssen04}. 
In order for a system to be in corresponding universality class is 
a fulfillment of four conditions: (i) a unique absorbing state, (ii) short-ranged interactions,
 (iii) a positive one-component order parameter,
 (iv) no additional property (symmetry, presence of additional slow variables, etc.)
 \cite{Janssen81,Grassberger82}.
As a prominent example DP was studied by diverse analytical and numerical methods \cite{HHL08}. Therefore it is
natural to consider DP when the main aim is to improve an existing method, what is part of our goal.
 Invaluable theoretical framework for an analysis of the scaling behavior is the renormalization group (RG) 
 method \cite{Zinn,Amit,Vasiliev}.
 In terms of RG flows of effective charges and accompanied existence of fixed points, 
divergent (power-law) behavior of various quantities can be naturally explained. Moreover, RG  
 enables us with different computational approaches for an approximate estimation of
 universal quantities in a controllable fashion. Famous scheme consists in
dimensional regularization augmented by so-called minimal subtraction (MS) scheme.
For DP this yields a perturbative
calculation in formally small parameter $\eps$, where $\eps$ is the deviation from the upper 
critical dimension $d_c = 4$. Regarding this line of reasoning
 an existing research has been mostly restricted to 
 the two-loop approximation of the perturbation theory \cite{Janssen04,Hinrichsen00}. 
 The main reason is that three-loop calculations put high demands on analytical methods \cite{lukas1,lukas2} and usually
 it is not possible to evaluate Feynman diagrams save by some efficient computational procedure. 
 
 Among the conditions of the DP universality class the item (iv) is 
 probably most relevant
 from the experimental point of view. In realistic setups impurities and defects
 are expected to cause deviations from DP universality class. 
 This is believed to be one of the reasons  why there are
not so many direct experimental realizations \cite{RRR03,TKCS07,Sano2016,LSAJAH16} of DP. 
 A study of deviations from the ideal situation could proceed in different routes and this 
 still constitutes a topic of an ongoing debate \cite{HHL08}.
A substantial effort has been made  in studying  a long-range interaction using
L{\'e}vy flights \cite{Hinrichsen06,Janssen99,Hinrichsen07}, effects
 of immunization \cite{Janssen04,Hinrichsen00}, or in the presence of spatially quenched
 disorder \cite{Janssen97}.
 Hence, as a further possible application of our methods, we analyze DP model in a presence of external
 velocity field that adds up to diffusion motion additional stirring effects.
 In this paper, we focus on DP in the presence
of advective velocity fluctuations, which are generated by means of Kraichnan model. 
%
%
Such problem was first proposed in the work~\cite{AntKap08}.
 There the model was analyzed using field-theoretic renormalization group to a leading one-loop approximation.
%
%

 Basic
idea of the model is to  assume that the velocity field can be imagined as a random Gaussian variable
 with prescribed  statistical properties \cite{Kra68,Ant99,FGV01}.
Despite obvious simplification in comparison to realistic flows, Kraichnan model is heavily
used in a fluid dynamics. His role is especially important in intermittency studies, because
it is one of the few models that allows an exact solution \cite{FGV01}.


The remainder of the paper is organized as follows. In 
 Sec.~\ref{sec:model}, we introduce a
coarse-grained formulation of the DP problem, and we give a brief description
of Kraichnan model for velocity fluctuations. Next we reformulate both models  
 into a field-theoretic language. In Sec.~\ref{sec:RG_analysis},  we present main steps of the
 perturbative RG analysis, and DP in presence of advection velocity field is renormalized to
 two-loop order.
 Sec.~\ref{sec:concl} is saved for a concluding summary. 

\section{Description of the Model} \label{sec:model}
The stochastic reaction-diffusion equation for a positive coarse-grain density of percolating particles 
$\psi = \psi (t, {\mx})$ has following form { {\cite{HHL08,Tauber2014}} }
\begin{equation}
  \partial_t \psi = D_0\left[\boldnabla^2 -\tau_0 \right]\psi - \frac{\lambda_0 D_0}{2} \psi^2
  +  \zeta \sqrt{\psi},
  \label{eq:dp-langevin}
\end{equation}
where $\partial_t = \partial / \partial t$, $\boldnabla^2 $ is the Laplace operator, $D_0$ is a
diffusion constant, $\lambda_0$ is a positive coupling constant and $\tau_0$ is a deviation from the 
threshold value of injected probability. It can be interpreted as a
 formal  analog of a deviation from critical temperature in static models \cite{Vasiliev}.
 Hereinafter the subscripts $0$ will always indicate an unrenormalized (bare) quantity.  The 
random Gaussian variable
 $\zeta (t, {\mx})$ {can be chosen in the following form \cite{Tauber2014}}
\begin{equation}
  \langle \zeta(t, {\mx}) \zeta(s, {\my}) \rangle = \lambda_0 D_0 \delta(t - s) 
  \delta^d({\mx}- {\my})
  \label{eq:dp-noise}
\end{equation}
with $d$-dimensional version of Dirac $\delta (x)$ function, i.e. 
 $\delta^d({\mx}- {\my})=\delta(x_1-y_1)\cdots\delta(x_d-y_d)$. The average $\langle\cdots\rangle$
 corresponds
 to a functional averaging over all noise realizations.
 
{Further step consists in an utilization of} 
 famous De Dominicis-Janssen formalism \cite{Janssen76,deDom76,Janssen79} that 
allows us to map the stochastic problem (\ref{eq:dp-langevin})-(\ref{eq:dp-noise}) 
 onto a field-theoretical model. Effectively one gets rid of noise variable, but on the other hand
 number of fields is doubled. This could be done in 
 a standard fashion \cite{Tauber2014,Vasiliev} and the resulting response functional 
 for the percolation process \cite{HHL08,Cardy80,Janssen04} reads
\begin{equation}
  \S^{DP} = \psid[-\partial_t + D_0 \nabla^2 - D_0 \tau_0] \psi +\frac{D_0 \lambda_0}{2}\left[ 
  \psid^2 \psi -\psid \psi^2\right],
  \label{eq:dp-action}
\end{equation}
where $\psid$ is an auxiliary Martin-Siggia-Rose response field, and the integration over the 
spatio-temporal arguments is implicitly assumed. For instance, first term on the right hand side 
actually stands for 
the expression $\psid\partial_t\psi = \int\dRM^d x\int\dRM t\,\psid(t,\mx)\partial_t \psi(t,\mx)$.

The percolation model is manifestly invariant \cite{HHL08} with respect to so-called
rapidity reversal symmetry
\begin{equation}
  \psi(t,\mx) \longleftrightarrow - \psid (-t,\mx).
  \label{eq:time-sym}
\end{equation}
This symmetry plays an important role in analysis of statistical quantities and among the practical
consequences is a reduction of number of independent critical indices \cite{HHL08}.

Main object of interest for
 the stochastic problem (\ref{eq:dp-langevin})-(\ref{eq:dp-noise}) are 
 statistical quantities, which  correspond to mean averages of expressions involving product of arbitrary
 number of fields. In field-theoretic formulation they are equivalently given as
functional averages over
the full set of fields with the ``weight'' functional $\exp (\S^{DP})$ and are known as
 correlation and response functions.
These functions are conveniently
 represented in the diagrammatic form of Feynman graphs \cite{Zinn,Vasiliev}. 

 Field-theoretic response functional (\ref{eq:dp-action}) is amenable to {a} standard 
 field-theoretic perturbative analysis. 
 Free part of the response functional yields just one bare propagator
$\langle \psi \psid \rangle_0$,
which takes the following form 
\begin{equation}
  \langle \psi \psid \rangle_0  = \theta (t) \exp[- D_0 (k^2 + \tau_0) t]  {,}
  \label{eq:DP_prop1}
\end{equation}
in time-momentum representation, whereas in frequency-momentum representation it takes the form
\begin{equation}
  \langle \psi \psid \rangle_0  =\frac{1}{-i \omega + D_0 (k^2+\tau_0)}.
  \label{eq:DP_prop2}
\end{equation}
Note that in \eqref{eq:DP_prop1} time flows from $\psid$ to $\psi$ field.
Non-linear terms in the response functional \eqref{eq:dp-action} give rise to two cubic vertices, whose
  vertex factors  \cite{Vasiliev} can {be} obtained using general formula
  \begin{equation}
  V_N(x_1,\ldots,x_N;\varphi) = 
  \frac{\delta^N \S_{\text{int}}[\varphi]}{\delta\varphi(x_1)\ldots\delta\varphi(x_N)},
  \quad
  \varphi \in\{ \psid,\psi,\mv \},
  \label{eq:ver_factor}
\end{equation}
{where} $\S_\text{int}$ is {a} non-linear part of the response functional.
  It is easy to {verify that from the action~\eqref{eq:dp-action}
  we obtain two vertex factors}
\begin{equation}
  V_{\psid^2 \psi} = - V_{\psid \psi^2} =  D_0 \lambda_0.
\end{equation}

 \begin{figure}[!ht]
     \centering
     \includegraphics[width=8cm]{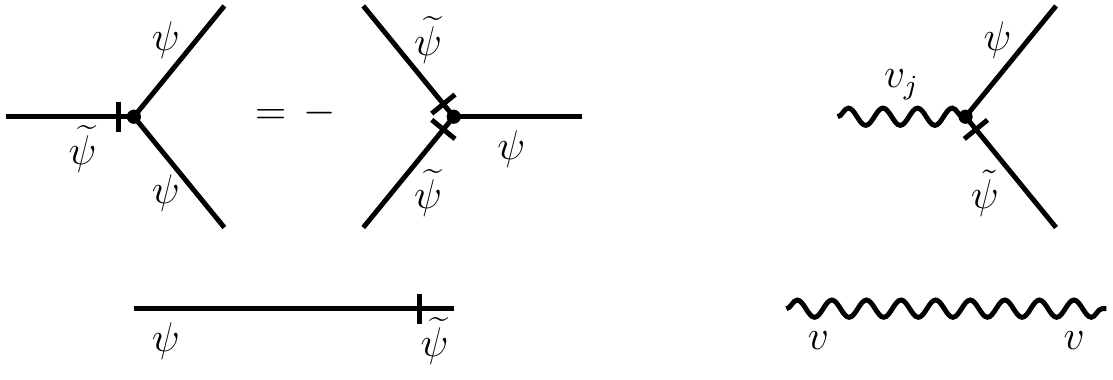}
     \caption{Diagrammatic representation of the bare
   propagators, the interaction vertices 
   describing an  ideal directed bond percolation process and  the influence of the
   advecting velocity field
   on the  order parameter fluctuations.}
     \label{fig:prop-ver}
 \end{figure}

In this paper, we analyze DP in a presence of additional velocity fluctuations. Basic underlying 
assumption
is that DP is advected, but does not {exert any backward} influence {on}
velocity field itself, i.e. it is an passive 
quantity \cite{Ant99}.
 Following {works}~\cite{Ant99,FGV01,Antonov06,AntKap08}
the turbulent mixing is described by velocity ensemble with prescribed statistics.
Inclusion of velocity field  
${\mv}(t,\mx) $ corresponds to a replacement 
\begin{equation}
  \partial_t \rightarrow \nabla_t = \partial_t + (\mv \cdot\boldnabla), 
  \label{eq:gal-invariance}
\end{equation}
 where $\nabla_t$ is the Lagrangian derivative \cite{Landau}. 
The velocity field will be assumed incompressible, i.e. {condition}
 $\boldnabla\cdot\mv = 0$ {is fulfilled}. According to Kraichnan suggestion we
 assume velocity field $\mv$ to be a Gaussian random variable with zero mean and prescribed correlator
\begin{align}
  \langle v_i (t, {\mx}), v_j (t', {\mx'}) \rangle & = \delta (t-t') D_{ij}
  ( {\mx} -  {\mx'}), 
  \label{eq:vel_correl1}
  \\
  D_{ij}  ( {\mx} -  {\mx'}) & = D_0 g_0 \int\limits_{k>m} \frac{\dRM^d k}
  {(2\pi)^d} P_{ij} (k) k^{-d-\xi}
  \eRM^{i{\mk}\cdot ({\mx} -{\mx'})}, \,\,  k \equiv |{\mk}|,
  \label{eq:vel_correl2}
\end{align}
where $P_{ij} (k) = \delta_{ij} - k_i k_j / k^2$ is transversal
projection operator, $g_0$ is small coupling constant, and the
cutoff $k=m$ provides an infrared regularization. Scaling exponent $\xi$ is related to
 a power-law behavior of energy spectrum \cite{Ant99}, and in RG approach plays a role
 of a formally small expansion parameter. 
Since the velocity fluctuations are governed by the Gaussian statistics, the
 corresponding averaging procedure is performed with the quadratic functional 
\begin{equation}
  \S_{\text{vel}}  = \frac{1}{2} 
  \int \dRM t_1 \int \dRM t_2 
  \int \dRM^d x_1 \int \dRM^d x_2  \,
  \mv_i(t_1,x_1) 
  D_{ij}^{-1}(t_1-t_2,\mx_1-\mx_2) \mv_j(t_2,\mx_2),
  \label{eq:vel_action}
\end{equation}
where $D_{ij}^{-1}$ is the kernel of the inverse linear operation {for the function
$D_{ij}  ( {\mx} -  {\mx'}) $} in (\ref{eq:vel_correl2}).

The full field theoretic model of the three fields 
$\varphi = \{\psid, \psi, {\mv} \}$ is described by the response functional with the following 
abbreviated form  
\begin{equation}
  \S = \psid[-\partial_t - ({\mv} \cdot \nabla ) + D_0 \nabla^2 - D_0 \tau_0] \psi +
  \frac{D_0 \lambda_0}{2}\left[
  \psid^2 \psi -\psid \psi^2\right] +\frac{1}{2} {\mv} D^{-1} {\mv}.
  \label{eq:total_action}
\end{equation}
 We see that
total response functional contains additional propagator $\langle v v\rangle_0$  and triple advection
vertex $\psid \psi v$. Its vertex factor~\eqref{eq:ver_factor} is
proportional to the momentum $i k_j$ of auxiliary field $\psid$. The graphical
representation of this vertex can be found in Fig.~\ref{fig:prop-ver}.

\section{Renormalization} \label{sec:RG_analysis}
A starting point of the perturbation theory is a free part of the response functional
given by  expression (\ref{eq:total_action}).
By graphical means, {it is} represented as lines in the Feynman diagrams, 
whereas the non-linear terms  correspond to vertices connected by these  lines.

For the calculation of the  RG constants  we employ dimensional regularization in the
combination with the {modified}  minimal subtraction ($\overline{\text{MS}}$) scheme \cite{Zinn}. 
It must be borne in mind
 that now we are dealing with double expansion approach \cite{HHL}. Therefore
 poles to the two-loop order that we encounter are of
 three types: either $1 / \eps$, $1 / \xi$ and $1/(\eps +\xi)$.
This simple picture pertains only to the lowest orders in a
perturbation scheme. In higher order terms, poles of
 general linear combinations in $\eps$ and $\xi$ are expected. 

The detailed examination of UV divergences is typically based on the analysis
of canonical dimensions \cite{Zinn,Vasiliev}. Dynamical
models have two scales, i.e. the canonical dimension of some quantity $Q$ is
described by two values, the momentum dimension $d_Q^k$
and frequency dimension $d_Q^{\omega}$. First, {it} is needed to introduce 
normalization conditions 
$d^k_{\mk}=-d^k_{\mx}=1$, $d^{\omega}_{\mk}=-d^{\omega}_{\mx}=0$, 
$d^k_{t}=d^k_{\omega}=0$,  $d^{\omega}_{\omega} = -d^{\omega}_{t}=1$.
The dimensions are then found from the requirement that each term of the response functional 
remains dimensionless (with respect to the momentum and
frequency dimensions separately). Further, the total canonical dimension 
$d_Q = d^k_Q + 2 d^{\omega}_Q$ plays the same role as momentum 
dimension in static models \cite{Vasiliev} and all canonical dimensions are
given in Tab.\ref{tab:canonical_dimension}.

\begin{table}[!ht]
\begin{center}
\caption{Canonical dimensions of the bare fields and bare parameters 
	  for the model}
\begin{tabular}{| c @{\hspace{0.5cm}} | c |  @{\hspace{0.5cm}}  c |  @{\hspace{0.5cm}}  c |  @{\hspace{0.5cm}}  c |
 @{\hspace{0.5cm}}  c |  @{\hspace{0.5cm}}  c | @{\hspace{0.5cm}}  c |  @{\hspace{0.5cm}} c | }
\hline
& \\[-2.5ex]
$Q$ & $\psi$ & $\psid$ & ${\mv}$ & $D_0$ & $\tau_0$ & $\lambda_0$ & $g_{0}$ & $u_{0}$  \\
\hline 
\hline
$d_Q^k$ & $d/2$ & $d/2$ & $-1$ & $-2$ & $2$ & $\eps/2$ & $\xi $ & $\eps$  \\
\hline
$d_Q^{\omega}$ & $0$ & $0$ & $1$ & $1$ & $0$ & $0$ & $0$ & $0$ \\
\hline
$d_Q$ & $d/2$ & $d/2$ & $1$ & $0$ & $2$ & $\eps/2$ & $\xi$ & $\eps$ \\
\hline
\end{tabular}
\label{tab:canonical_dimension} 
\end{center}
\end{table}

Crucial objects in an analysis of translationally invariant theories are $1$-irreducible Green function 
 $\Gamma = \langle \varphi \dots \varphi\rangle_\text{1-ir}$, which
are derived from connected Green function by an appropriate Legendre transformation \cite{Vasiliev}. 
Canonical dimension of $\Gamma$ {is given by the relation}
\begin{equation}
  d_{\Gamma} = {d +2 - n_{\varphi}d_{\varphi},}
\end{equation}
where $n_{\varphi}= \{n_{\psid},n_{\psi}, n_{\mv }\}$
represent the number of fields appearing in
the function $\Gamma$. Superficial UV
divergences can be generated only in those Green {functions} for which $d_\Gamma$ is a 
nonnegative integer. For the pure DP model \cite{Janssen04} {UV divergences are present} 
in the following $1$-irreducible functions:
$\langle \psid \psi \rangle_\text{1-ir}$, 
$\langle \psid \psi \psi \rangle_\text{1-ir}$, $\langle \psid \psid \psi \rangle_\text{1-ir}$ and
their corresponding counterterms are already present in response functional (\ref{eq:dp-action}). 

By direct inspection of the Feynman diagrams one can 
observe that the real expansion parameter in perturbation theory is 
 $\lambda_0^2$ instead of $\lambda_0$. 
 This could be easily seen by a direct examination of 
Feynman diagrams {and can be regarded as} a direct consequence of the rapidity-reversal 
 {symmetry~\eqref{eq:time-sym}.}
 For this reason it is convenient to introduce a new charge 
\begin{equation}
  u_0 =\lambda^2_0,
  \label{eq:new_u}
\end{equation} 
where $u_0$ has canonical dimension $4-d=\eps$. The perturbative calculation is then made
 in terms of $u_0$.

The total renormalized response functional for DP in a presence of advecting velocity
fluctuations takes the following form
\begin{equation}
  \S_R  = \psid[- Z_1 \partial_t + Z_2 D \nabla^2 - Z_3 D \tau] \psi +
  \frac{Z_4 D \lambda \mu^{\eps/2}}{2}[\psid^2 \psi - \psid \psi^2] 
    - Z_1 \psid({\mv} \cdot \boldnabla ) \psi +
  \frac{1}{2} {\mv} D^{-1} {\mv}{, }
  \label{eq:action-dp-r}
\end{equation}
where $\mu$ is renormalization mass \cite{Zinn,Vasiliev}. The model is assumed to be in
 a scaling
 region, which is obtained for $\tau_0$ close enough to its critical value. 
  In order to preserve the
Galilean invariance \cite{Vasiliev} the advection term and the term containing
temporal derivative \eqref{eq:gal-invariance} {have to be renormalized by} the same 
 {renormalization} constant.
In addition, the last term in the {action} \eqref{eq:action-dp-r} is
not renormalized at all due to 
%
%
a passive nature of the advecting scalar field $\psi$. This ensures
nonexistence of nontrivial Feynman diagrams for velocity propagator \cite{Ant99}.
%
%
The amplitude factor is then expressed as
\begin{equation}
  g_0 D_0  = g D \mu^{\xi},
\end{equation}
and the renormalization constants are related as follows
\begin{equation}
  Z_g Z_D = 1, \hspace{1cm}    Z_g^{-1} = Z_D = Z_2 Z_1^{-1}.
\end{equation}

The renormalized response functional can {also be} obtained
 by renormalization of fields and parameters
\begin{align}
  &\psi \rightarrow \psi Z_{\psi}, & \psid &\rightarrow \psid Z_{\psid}, & v &\rightarrow Z_v v,
   &\tau_0 & =  \tau Z_{\tau} + \tau_c, \nonumber \\
  & D_0 = D Z_D, & \lambda_0 & =  \lambda \mu^{\eps/2}Z_{\lambda}, &u_0 & =  u \mu^{\eps} Z_{u}, 
  & g_0 & =  g \mu^{\xi} Z_g, 
\end{align}
where we have symbolically expressed needed renormalization of fields.
Relations among the renormalization constants take following form 
\begin{equation}
  Z_1 = Z_{\psi} Z_{\psid}, \quad Z_2 = Z_D Z_{\psi} Z_{\psid},\quad
  Z_3 = Z_{\tau} Z_D Z_{\psi} Z_{\psid}, \quad Z_4 = Z_{u}^{1/2} Z_D Z_{\psi}^2 Z_{\psid}.
  \label{eq:rg-constants}
\end{equation}
On the other hand, the renormalization constants {for} fields and parameters can be expressed by the inverse
 formulas
\begin{equation}
  Z_{\psi} = Z_{\psid} = Z_{1}^{1/2}, \quad Z_D = Z_{2} Z_1^{-1}, \quad
  Z_{\tau} = Z_3 Z_{2}^{-1}, \quad Z_{u} = Z_4^2 Z_{2}^{-2} Z_{1}^{-1}.
  \label{eq:RG_constants}
\end{equation}
 {As has been pointed out, passive nature of the problem}
 ensures that the renormalization {constant} for velocity field $v$ is simply $Z_v=1$.

Next, we briefly show derivation of RG equation \cite{Zinn,Vasiliev} needed for an overall analysis of scaling 
 behavior. The basic idea is corroborated by the claim that
 {renormalized} Green functions differ from unrenormalized ones by rescaling of the fields 
and choice of parameters. The fundamental relation
$\S_R (\varphi, e, \mu) = \S (\varphi, e_0)$ between response functional (\ref{eq:dp-action}) and
(\ref{eq:action-dp-r}) leads directly to the formula
\begin{equation}
  \Gamma_R \left(e,\mu, \dots \right) = Z_{\varphi}^{n_{\varphi}} \Gamma \left(e_0,  \dots \right),
\end{equation}
where $e_0 = \{ D_0, \tau_0, u_0, g_0\}$ is full set of bare parameters and  $e = \{ D, \tau, u, g\}$
are their renormalized counterparts,
the ellipsis stands for the other arguments (time/frequency, coordinates/momenta of appearing fields).
Further, {we} denote $\mu$-derivatives at fixed bare parameters by 
$\widetilde{D}_{\mu} = \mu \partial_{\mu}$. The equation $\widetilde{D}_\mu \Gamma = 0$ then
 yields basic RG differential
equation for the renormalized Green function $\Gamma_R$
\begin{equation*}
  \left[ D_{RG} - n_{\varphi} \gamma_{\varphi}\right]\Gamma_R = 0,
\end{equation*}
where $D_{RG}$ is the operation $\widetilde{D}_{\mu}$ expressed in the renormalized quantities 
\begin{equation}
  D_{RG} \equiv \mu \partial_{\mu} + \beta_{u}\partial_{u} + 
  \beta_{g}\partial_{g}- \tau \gamma_{\tau} \partial_{\tau} 
  - D \gamma_D \partial_D.  
\end{equation}
The anomalous dimension $\gamma_F$ for any quantity $F$ is given by the relation
\begin{equation}
  \gamma_F = \widetilde{D}_{\mu} \ln Z_F, \quad F \in \{\psi,\psid, D, \tau, u, g\}.
  \label{eq:gamma1}
\end{equation}
 $\beta$-functions for the dimensionless couplings $u$ and $g$ {are}
\begin{equation}
  \beta_u \equiv \widetilde{D}_{\mu} u = u \left[-\eps -\gamma_u\right], \quad
  \beta_g \equiv \widetilde{D}_{\mu} g = g \left[-\xi -\gamma_g\right]. 
  \label{eq:beta}
\end{equation}
 The relations among the anomalous dimensions fields, parameters and $\gamma_i$ take the form 
\begin{equation}
  \gamma_1 = 2 \gamma_{\psi}, \quad
  \gamma_2 = 2 \gamma_{\psi} + \gamma_D, \quad
  \gamma_3 = 2\gamma_{\psi} +\gamma_D +\gamma_{\tau}, \quad 
  \gamma_4 = 3 \gamma_{\psi} + \gamma_D + \frac{1}{2}  \gamma_{u}.
  \label{eq:gamma2}
\end{equation}
Let us note that in what follows our main aim is to analyze phase structure of the theory.
To this end the calculation of $\gamma_3$ is not needed and therefore we do not
consider it here. 

%
%
Large-scale (macroscopic) regimes of a given renormalizable field theoretic model are associated
with IR attractive fixed 
points of the corresponding RG equations \cite{Zinn,Vasiliev}.  A fixed point (FP) is defined as a such point 
$ (g^{*}, u^{*})$ for which $\beta$
functions $\beta_g$ and $\beta_u$ simultaneously vanish, i.e. 
\begin{equation}
  \beta_{g}(g^{*},u^*) = \beta_{u}(g^{*},u^*) = 0.
  \label{eq:zero_beta}
\end{equation}
The IR stability of a fixed point is then determined by the matrix of first derivatives of $\beta$ functions
\begin{equation}
  \Omega_{ij} = \frac{\partial \beta_i}{\partial g_j}\biggl|_{*},\quad i,j\in\{g,u\}.
  \label{eq:omega}
\end{equation}
Asterisk in this equation indicates  a corresponding fixed point value.
The IR-asymptotic {behavior} is governed by IR-stable fixed points, for which real parts of all 
eigenvalues of matrix (\ref{eq:omega}) are positive.
%
%

The renormalization constants absorb all divergences at $\eps$, $\xi \rightarrow 0$ and the
renormalized functions {are}
finite for $\eps$, $\xi = 0$. For the RG calculation the modified minimal subtraction  
scheme $\overline{\text{MS}}$ has been chosen
\cite{Collins}. Difference between $\overline{\text{MS}}$ and minimal subtraction 
 is that factor $S_d/(2\pi)^d$ is
not expanded in {$\varepsilon=4-d$}. In the present case the RG constants attain a general form
\begin{equation}
  Z_i=1+\text{ pole terms in }\eps, \xi \text{ and their linear combinations}.
\end{equation}
Coefficients in this
expansion depend on two coupling constant $u$ and $g$. The coupling {constants} have been rescaled by a 
 convenient constant factor
\begin{equation}
  \frac{g S_d}{2 (2 \pi)^d} \rightarrow g,\quad
  \frac{u S_d}{2 (2 \pi)^d} \rightarrow u.
  \label{eq:new_charge}
\end{equation}
Dyson equation for the function $\langle \psid \psi \rangle$ {reads}
\begin{equation}
  \Gamma^R_{\psid \psi} = i \omega Z_1 + D p^2 Z_2+ D \tau Z_3  - {\sum}_{\psid\psi}^\text{adv} 
  - {\sum}_{\psid\psi}^\text{perc},
  \label{eq:1PIpsi}
\end{equation}
where $\sum^\text{perc}$ {contains} only contributions arising {solely} from
the pure DP process, whereas $\sum_{\psid\psi}^\text{adv}$
corresponds to diagrams containing {velocity} 
propagator $\langle v v \rangle$. {Two-loop} 1-irreducible Feynman diagrams with nonzero contribution to Green 
function~\eqref{eq:1PIpsi} are the following 
\begin{align}
  {\sum}_{\psid\psi}^\text{adv}
  & =
  \raisebox{-16pt}{\includegraphics[width=2.5cm]{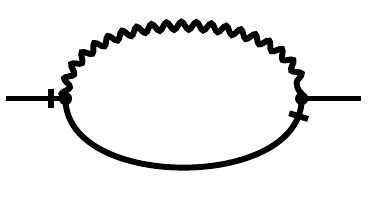}} +
  \raisebox{-2pt}{\includegraphics[width=3cm]{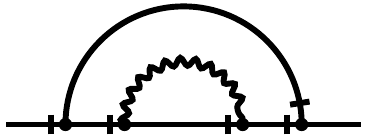}} +
  \frac{1}{2} \raisebox{-18pt}{\includegraphics[width=3cm]{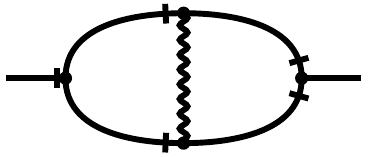}},
  \label{eq:pert_exp1} \\
  {\sum}^\text{perc}_{\psid\psi}
  & = \frac{1}{2}
  \raisebox{-16pt}{\includegraphics[width=2.5cm]{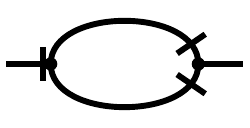}} +
  \frac{1}{2}
  \raisebox{-2pt}{\includegraphics[width=3cm]{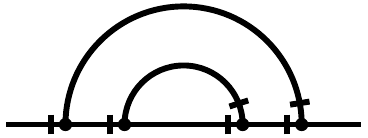}} +  
  \raisebox{-18pt}{\includegraphics[width=3cm]{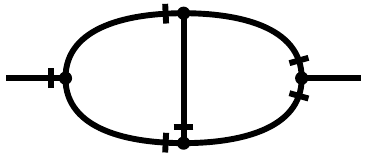}}.
  \label{eq:pert_exp2}
\end{align}

The counterterms in the $\overline{\text{MS}}$ scheme are polynomials in 
IR regulators. Furthermore, the RG constants $Z_i;i=1,2,3,4$ could not {depend} on the
choice of the IR regularization \cite{Zinn,Vasiliev}. In the case of the pure DP process 
$\tau$ is the IR regulator
\cite{Janssen04}. From the practical point of 
view, it is advantageous to set $\tau = 0$ in the response functional (propagator 
$\langle \psi \psid \rangle$ and cut off the momentum integrals at
$k = m$ (by dimension $\tau \sim m^2$). 

The renormalization constants can be expressed as follows
\begin{align}
  Z_1 &= 1+\frac{u}{4\varepsilon}+\frac{u^2}{32\varepsilon}  \bigg[ \frac{7}{\varepsilon}-
  3+\frac{9}{2}\ln \frac{4}{3} \bigg]+
  \frac{ug}{16}\bigg[\frac{6}{\varepsilon(\varepsilon+\xi)} +\frac{1}{\varepsilon+\xi}\bigg],\\
  Z_2 &= 1+\frac{u}{8\varepsilon} +g\frac{3}{4\xi} +\frac{u^2}{128\varepsilon} \bigg[ 
  \frac{13}{\varepsilon}-\frac{31}{4} +
  \frac{35}{2}\ln \frac{4}{3} \bigg] + 
  \frac{ug}{128} \bigg[-\frac{24}{\xi\varepsilon} \nonumber \\
  &+\frac{36}{\varepsilon(\varepsilon+\xi)}+\frac{2}{\xi} + \frac{9}{\varepsilon+\xi} \bigg].
\end{align}
Let us illustrate a calculation of anomalous dimension $\gamma_1$ using~\eqref{eq:gamma1}.
 As a first step we derive approximate relation
\begin{equation*}
  \ln Z_1 \approx\frac{ u}{4 \varepsilon }+ \frac{g u}{8 (\xi +\varepsilon )}\Big( \frac{1}{ \varepsilon  } +\frac{1}{2} \Big)+\frac{3 u^2}{16 \varepsilon }\Big( \frac{1}{\varepsilon}
  -\frac{1}{2 }+\frac{3 }{4 \varepsilon }\ln \frac{4}{3} \Big)+\mathcal{O}(u^3)+\mathcal{O}(g u^2)+\mathcal{O}(g^2 u),
\end{equation*}
where the last three terms stand for higher order terms that are neglected in what follows. 
Next, we need a formula $ (\beta_u\partial_u + \beta_g \partial_g)$, which can be approximated as follows
\begin{equation*}
  u(-\varepsilon+\gamma^{(1)}_u)\partial_u + g(-\xi+\gamma^{(1)}_g) \partial_g
  \approx u\Big(-\varepsilon+\frac{3 u}{2}+ \frac{3 g}{2}\Big)\partial_u + 
  g \Big(-\xi+\frac{u}{8} +\frac{3 g}{4}\Big)\partial_g, 
\end{equation*}
  where $\gamma^{(1)}_u$ and $\gamma^{(1)}_g$ are appropriate gamma functions up to the first 
  order in perturbation theory.  
Finally, we get
\begin{equation}
  \gamma_1=   -\frac{u}{4}+\frac{u^2}{32}\Big( 6-9\ln\frac{4}{3} \Big) -\frac{ug}{16},
\end{equation}
where terms proportional to {$ ug\xi/(\varepsilon(\varepsilon+\xi))$} and 
{$u^2/(\varepsilon+\xi)$} drop out. In a similar fashion anomalous dimensions $\gamma_2$ and 
$\gamma_4$ can be calculated. Using them we derive in a straightforward way
anomalous dimensions for $\psi$ field and diffusion constant $D$
\begin{align}
  \gamma_{\psi}&= -\frac{u}{8}+\frac{u^2}{64}\Big( 6-9\ln\frac{4}{3} \Big) -\frac{ug}{32},
  \label{eq:gammap}
  \\
  \gamma_D&= - \gamma_g= \frac{1}{8}u+\frac{3}{4}g-\frac{u^2}{256}\Big( 17 -
  2\ln\frac{4}{3} \Big) - \frac{ug}{128}\Big( 3 + 2\zeta \Big),  
  \label{eq:gammaD}
\end{align}
where the ratio $\eps/\xi = \zeta$ is a finite quantity \cite{HHL}. 
In the models with two regulators such as $\eps$ and $\xi$, it is usually
assumed that they are of the same order $\eps = \mathcal{O} (\xi)$. For $g = 0$ (DP model without
advection interactions) expressions \eqref{eq:gammap}-\eqref{eq:gammaD} coincide
with known two-loop results \cite{Janssen04,Janssen81}. 
Further, Feynman diagrams contributing to RG constant $Z_4$ are the following
\begin{align}
  {\sum}_{\psid\psi\psi}^\text{adv} & = 
  2 \raisebox{-27pt}{\includegraphics[width=3cm]{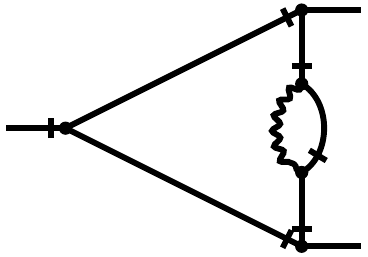}} +
  2 \raisebox{-27pt}{\includegraphics[width=3cm]{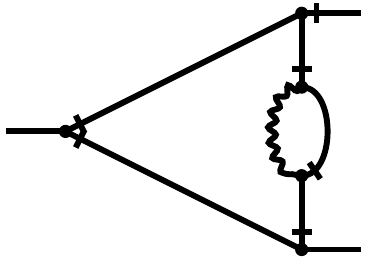}} + 
  2 \raisebox{-27pt}{\includegraphics[width=3cm]{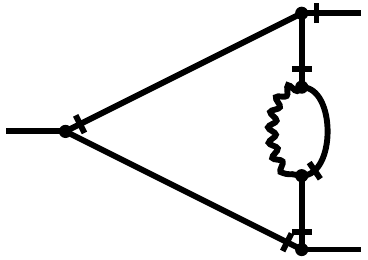}} 
  + \nonumber \\
   & + 2 \raisebox{-27pt}{\includegraphics[width=3cm]{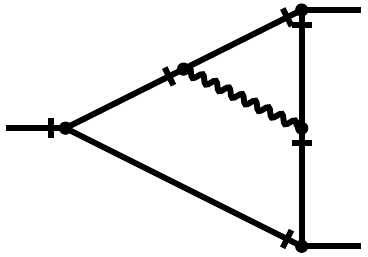}} +
   2 \raisebox{-27pt}{\includegraphics[width=3cm]{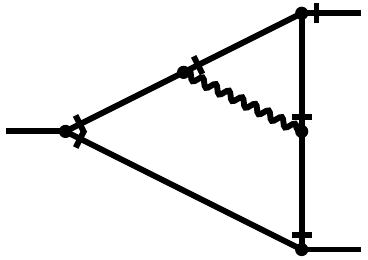}},
\end{align}
where the factor two in front of the diagrams accounts for {an} internal symmetry of the graph, i.e.
 it expresses number of ways for drawing a given topological configuration. Altogether, the 
 final expression for RG constant $Z_4$ reads
\begin{equation}
  Z_{4}=1+\frac{u}{\varepsilon} +\frac{u^2}{16\varepsilon} \bigg[ \frac{20}{\varepsilon}-
  {7} \bigg]+ \frac{ug}{4}\bigg[ 
  \frac{6}{\varepsilon(\varepsilon+\xi)} +\frac{1}{\varepsilon+\xi} \bigg].
\end{equation}
Then, using~\eqref{eq:RG_constants} we can calculate anomalous dimension of the charge $u$
\begin{equation}
  \gamma_u =-\frac{3}{2}u-\frac{3}{2}g+ \frac{u^2}{128}\Big(
  {169} + 106\ln\frac{4}{3} \Big) + 
  \frac{ug}{64} \Big( {2\zeta} - 17  \Big).
  \label{eq:anom_u}
\end{equation}
This expression finalizes two-loop perturbative renormalization of the model. Using explicit
information about $Z_1,Z_2$ and $Z_4$ allows one to determine fixed {points'} structure and thus
find scaling regimes. 

There are no other two-loop diagrams of an order $g^2$ due to a fact that in pure Kraichnan model
for passive admixture all higher order corrections vanish \cite{Ant99}. Knowledge
of anomalous dimensions $\gamma_g$ and $\gamma_u$ along with $\beta$-functions~\eqref{eq:beta} 
allows us to perform a full two-loop analysis of fixed points' structure.
First and foremost there is a trivial or Gaussian fixed point FPI with coordinates
	\begin{equation}
	  u^*=  0,\quad g^*= 0. 
	\end{equation}
This corresponds to a fixed point (FP) with  irrelevant both DP interactions and advection process, 
and standard perturbation theory is applicable. As expected, this regime is IR stable  in the region
	\begin{equation}
	  \varepsilon<0, \quad \xi<0.
	\end{equation}
The former condition ensures that we are above the upper critical dimension $d_c=4$.	

Next, there is a FP point  FPII that corresponds to a pure DP process without advection. Its
coordinates are		
	\begin{equation}
	  u^*= \frac{2 \eps }{3}+ \frac{1}{432} \eps ^2 \bigg(169+106 \ln \frac{4}{3}\bigg), \quad g^*= 0.
	\end{equation}
Condition $g^*=0$ ensures that velocity propagator is effectively {irrelevant}. 
%
%
Eigenvalues of the matrix~\eqref{eq:omega} are
\begin{align}
   \lambda_1 & = \eps -\frac{\eps^2}{288} \left(
   169+106\ln \frac{4}{3}   
   \right) ,\quad
   \lambda_2  = \frac{\eps}{12} -\xi + \frac{\eps^2}{3456}\left(
   67+108\ln \frac{4}{3}
   \right)
   .
\end{align}
The first eigenvalue $\lambda_1$ agrees with a known two-loop result \cite{Janssen04}.
From an inspection of second eigenvalue $\lambda_2$ we observe that $\xi$ is restricted
by a parabolic function of $\eps$.
%
%

Coordinates of third FPIII are
	\begin{equation}
	  u^*= 0,\quad	g^*= \frac{4 \xi }{3},
	\end{equation}
and it is IR stable in the region	
	\begin{equation}
	  \xi > 0 ,\quad \xi > \frac{ \eps }{2}.
	\end{equation}
This FP corresponds to the pure advection process for which DP non-linearities are irrelevant.	

\begin{figure}[h!]
	\centering
	\includegraphics[width=6.cm]{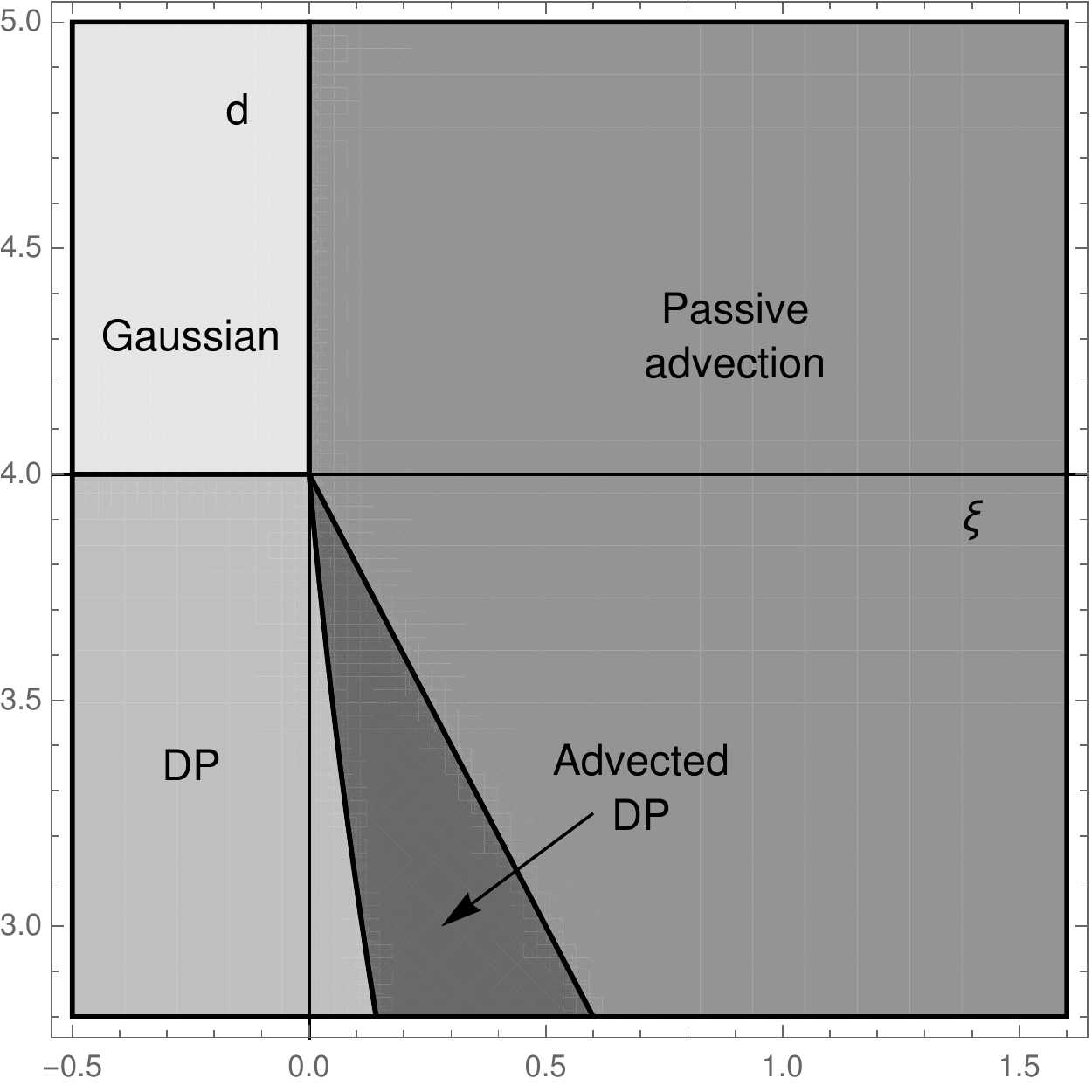} 
	 \caption{Regions of stability for the scaling regimes for DP process in a presence
	 of velocity fluctuations.
      The borders between the regions are depicted with the bold lines.}
      \label{fig:phase_diag}
\end{figure}

Last FPIV is most interesting, because both non-linearities are IR relevant. Mutual interplay between
DP and advection processes give rise to non-trivial behavior. The coordinates are
	\begin{align}	
	  u^* & = \frac{4}{5}  (\eps - 2\xi)+ ( 2\xi - \eps ) 
	  \bigg[-\varepsilon\Big( \frac{238}{375}+\frac{54}{125}\ln
	  \frac{4}{3}  \Big) +\xi\Big(\frac{192}{125}+\frac{108}{125}\ln \frac{4}{3}    \Big) \bigg],\\
	  g^* & =\frac{2}{15} (-\eps + 12\xi)+ \frac{ 2\xi-\eps}{75}
	  \bigg[\eps \Big( \frac{59}{15}+\frac{\zeta}{6}
	  +\frac{59}{10}\ln \frac{4}{3}  \Big) 
	  -\xi\Big(\frac{137}{10}+ {2\zeta}+\frac{59}{5}\ln \frac{4}{3}    \Big) \bigg],	
	\end{align}
and this FP is stable in region	
	\begin{equation}
          \xi<\frac{\eps}{2},\quad \xi> 0.0833 \eps + 0.0286 \eps^2,
	\end{equation}
where the second inequality is obtained by numerical calculation with error smaller then $10^{-4}$. 
%
%
 While the second-loop approximation does not qualitatively change one-loop results \cite{AntKap08}, 
	we see that now the boundaries between the regions
of stability are described not by lines in contrast to the one-loop result, but rather by parabolic curves.
%
%
For a better visual aid,
stability regions in $(d,\xi)-$plane {are} depicted in Fig.~\ref{fig:phase_diag}.
Note that only boundary between fixed points FPII and FPIV becomes parabolic due to two-loop corrections.
{ \section{Conclusion} \label{sec:concl} }
In this paper,  we have investigated an effect of
{incompressible} velocity fluctuations on the directed percolation process. 
Main points of field-theoretic formulation with
inclusion of the  advecting velocity field have been shown together with a 
renormalization group analysis. 

We have established that depending on the values of a spatial dimension $ d = 4-\eps $ and scaling
exponent $ \xi $, describing scaling properties of velocity fluctuations, the model exhibits four
distinct universality classes. They correspond to: the Gaussian (free) fixed point,
 a directed percolation without advection, a passive scalar advection, and fully non-trivial
 regime, in which both percolation and advection interactions are relevant. 
All relevant quantities, such as fixed points coordinates, regions of stability and anomalous
dimensions $\gamma_\psi,\gamma_D$ and $\gamma_u$ have been calculated up to two-loop approximation.
%
%
Despite obvious technical difficulties related to two-loop calculations, main physical consequences
are in accordance with the previous one-loop result~\cite{AntKap08}.
%
%

 The purpose of this paper was two-fold. First, our aim was to improve existing results 
 in non-equilibrium physics, which are mostly restricted to one-loop order.  Second, this article may
 be considered as a first  step in more challenging
 attempt, which would correspond to velocity field generated by some microscopic model such as stochastic
 Navier-Stokes equation {in two-loop approximation}. 
\section*{Acknowledgment}
The authors thank to Loran~Ts.~Adzhemyan, Juha~Honkonen and Nikolay~M.~Gulitskiy for  illuminating and fruitful discussions.
The work was supported by VEGA grant No. 1/0345/17 of the Ministry of
Education, Science, Research and Sport of the Slovak Republic, and the grant of the Slovak
Research and Development Agency under the contract No. APVV-16-0186.

\section*{Appendix A: Explicit calculation of Feynman diagram}
In this section we  present main steps of a typical calculation of a divergent part of a Feynman diagram.
  Let us consider second diagram in an expansion~\eqref{eq:pert_exp1} for 
   1-irreducible Green function $\langle\psid\psi\rangle_\text{1-ir}$.  
   We choose external {parameters} $ p =(\mpp,\Omega) $ to enter the diagram
   from the left. There are internal {variables} $ k=(\mk,\omega_k) $ and
   $ q=(\mq,\omega_q) $ {over} which we have to integrate.
   Using the standard Feynman diagrammatic technique, we construct the following algebraic expression for
   the diagram  
  \begin{align}
    		&\raisebox{-2.6ex}{\includegraphics[width=8.cm]{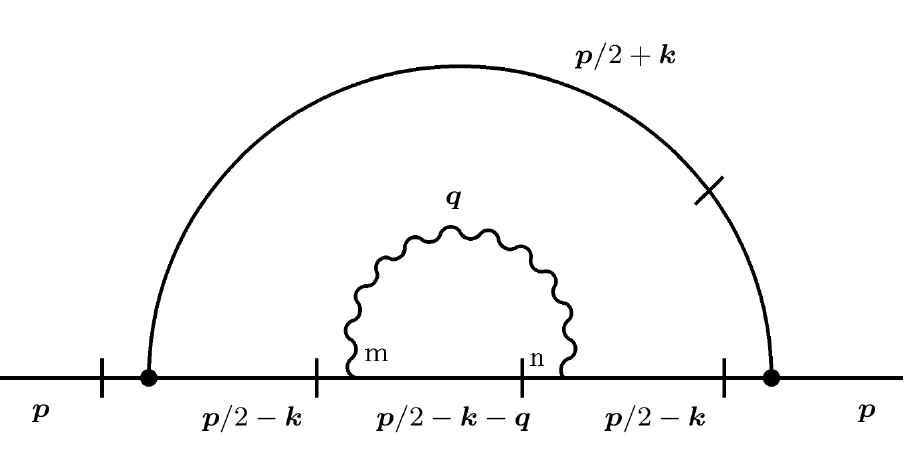}} \label{eq:pekny_diag} \\
		&=\frac{1}{(2\pi)^{2d+2}}\int\!\dRM^d  k\! \int\!\dRM^d q\int\!\dRM
		\omega_q\!\int\dRM \omega_k \frac{D_0g_0(\lambda_0D_0)
		(-\lambda_0D_0)}{ 2 \left[ D_0 \left(\left(\frac{p}{2}-k\right)^2+\tau_0 \right)
		-i \left(\frac{\Omega }{2}-\omega _k\right) \right]^2 } \nonumber \\
		&\times\frac{P_{mn}(q)q^{-d-\xi}[-i (p/2-k)_m][-i(p/2-k-q)_n] }{\left[D_0 
		\left(\left(-k+\frac{p}{2}-q\right)^2+\tau_0 \right)-i \left(-\omega_k
		-\omega _q+\frac{\Omega }{2}\right)\right]} 
		\frac{1}{\left[D_0 \left(\left(k+\frac{p}{2}\right)^2+\tau_0 \right)-
		i \left(\omega _k+\frac{\Omega }{2}\right)\right]}, \nonumber   
  \end{align}
where indices $m,n$ denote vector components of a velocity propagator~\eqref{eq:vel_correl1}.
Using Cauchy integral formula integration over frequency variables $\omega_k$ and $\omega_q$
  is readily performed. 
  In addition, a straightforward simplification of the tensor structure is possible and finally
  we arrive at the following expression
\begin{equation}
  \!\frac{ D_0^3 g_0\lambda_0^2}{8(2\pi)^{2d}}\!\!\int\!\!\dRM^d  k\! \int\!\! \frac{\dRM^d q }{ q^{d+\xi+2}}
  \frac{ 4k^2q^2 + 4 (\mk\cdot \mq) (\mpp\cdot \mq)- 4 (\mpp\cdot \mk)q^2-
  4 (\mk\cdot \mq)^2+p^2q^2-(\mpp\cdot \mq)^2  }{ 
  \left[-i\Omega + D_0 \left(\left(\frac{\mpp}{2}-\mk\right)^2+\left(\mk+\frac{\mpp}{2}\right)^2+2
  \tau_0 \right)\right]^2}.
  \label{eq:pert_exp3}
\end{equation}
We are interested in UV divergent parts, which are known to be proportional to external frequency $\Omega$,
 square of external momentum $\mpp$ and mass term $\tau$. We expand~\eqref{eq:pert_exp3} in a
 Taylor series and make a following substitution
$\cos \theta \rightarrow z$ for a scalar product $(\mk\cdot\mq)=k q\cos \theta$ between
 internal momenta $\mq$ and $\mk$. In other words
$\theta$ is the angle between the {vectors} $\mk$ and $\mq$. 
\begin{align}
  & -\lambda_0^2 g_0\frac{S_dS_{d-1}}{8(2\pi)^{2d}}
  \int\!\!\frac{\dRM  k}{k^{1+\varepsilon}}\! \int\!\!\frac{\dRM q}{q^{1+\xi}}
  \int\limits_{-1}^{1}\! \dRM z(1-z^2)^{\frac{d-1}{2}} 
  \biggl[ i\Omega
  +2\tau_0D_0
  +p^2D_0 \frac{ (-1 + d (-1 + 2 z^2))}{4d(1-z^2)}
  \biggl],   
\end{align}
where  $ S_{d} = 2\pi^{d/2}/\Gamma(d/2) $ is the  surface of a $ d $-dimensional sphere, and 
 for a calculation of $\mpp^2${-}term following formula
\begin{align}
  \int \dRM^{d}k\,  k_{i}k_{j} f(k^{2}) = \frac{1}{d} \int \dRM^{d} k \, k^{2} f(k^{2}) \color{red}{,}
  \label{eq:nConst_Int1}
\end{align}
was used~\cite{Vasiliev}. Using a definition of charge $u$ from Eq.~\eqref{eq:new_u}, substitution~\eqref{eq:new_charge}
and a relation $d=4-\eps$, we finally
obtain a expression for the UV divergent part of a diagram~\eqref{eq:pekny_diag}
\begin{equation}
  ug\Big(\frac{\mu}{m}\Big)^{\eps + \xi} 
  \biggl[    
  4i\Omega
     +8\tau D 
  - Dp^2 
  \biggl] \frac{12-\eps}{128\eps\xi}.
\end{equation}



\begin{thebibliography}{99}    
  \bibitem{Zia95} B. Schmittmann, R. K. P. Zia, {\it Phase transitions and critical phenomena}, 17 (1995).
  \bibitem{HHL08} M.~Henkel, H.~Hinrichsen, S.~L{\"u}beck, 
     {\it Non-equilibrium phase transitions: Volume 1 – Absorbing phase transitions}, 
     (Springer, Dordrecht, 2008).
  \bibitem{Tauber2014} U. C.~T{\"a}uber, {\it Critical Dynamics: A Field Theory Approach to Equilibrium and Non-Equilibrium Scaling Behavior}, Cambridge University Press, Cambridge (2014).
  \bibitem{Krapivsky} P. L.~Krapivsky and S. Redner and E. Ben-Naim, {\it A Kinetic View of Statistical Physics}, Cambridge University Press, Cambridge (2010).
  \bibitem{Zinn} J. Zinn-Justin, {\it Quantum Field Theory and Critical Phenomena}, Claredon Press 1989, 4th  edn. Oxfod University Press (2002).
  \bibitem{Amit} D.~J.~Amit, {\it Field Theory, the Renormalization Group, and Critical Phenomena}, World Scientific, Singapore (1984).
  \bibitem{Stauffer} D.~Stauffer, A.~Aharony, {\it Introduction to Percolation Theory}
     (Taylor and Francis, London, 1992).
  \bibitem{Cardy80} J. L. Cardy and R. L. Sugar, J. Phys. A {\bf 13}, L423 (1980).
  \bibitem{Odor04} G. \'Odor, Reviews of Modern Physics {\bf 76},  663 (2004).
  \bibitem{Janssen04} H. K. Janssen and U. C. T\"{a}uber, \emph{Ann. Phys.} {\bf 315}, 147 (2004).
  \bibitem{Janssen81} H. Janssen, Z. Phys. B: Cond. Mat.  {\bf 42}, 151 (1981).
  \bibitem{Grassberger82} P. Grassberger, Z. Physik B: Cond. Mat. {\bf 47},  365 (1982).
  \bibitem{Vasiliev} A. N. Vasil'ev, {\it The Field Theoretic Renormalization Group  in Critical Behavior Theory and Stochastic Dynamics}, [in Russian], PIYaF, St. Petersburg (1998); English trans., Chapman and Hall/CRC, Boca Raton, Fla (2004).
  \bibitem{Hinrichsen00} H. Hinrichsen, Adv. Phys. {\bf 49}, 815–958 (2000).
    \bibitem{lukas1} L. Ts. Adzhemyan, M. Hnati\v{c}, M. V. Kompaniets, T. Lu\v{c}ivjansk\'y, L. Mi\v{z}i\v{s}in,
                    EPJ Web of Conf. {\bf 108}, 02005 (2016).  
  \bibitem{lukas2} L. Ts. Adzhemyan, M. Hnati\v{c}, M. V. Kompaniets, T. Lu\v{c}ivjansk\'y, L. Mi\v{z}i\v{s}in,
                    EPJ Web of Conf. {\bf 173}, 02001 (2018). 
  \bibitem{RRR03} P.~Rupp, R.~Richter, I.~Rehberg, Phys. Rev. E {\bf 67}, 036209 (2003).
  \bibitem{TKCS07} K. A. Takeuchi, M. Kuroda, H. Chaté, M. Sano, {Phys. Rev. Lett.} {\bf 99}, 234503 (2007).
  \bibitem{Sano2016} M. Sano, K. Tamai, {Nat. Phys.} {\bf 12}, 249 (2016).
  \bibitem{LSAJAH16} G. Lemoult, L. Shi, K. Avila, SV Jalikop, M. Avila, and B. Hof, {Nat. Phys.} {\bf 12},
  254 (2016).
  \bibitem{Hinrichsen06} H.~Hinrichsen, Physica A {\bf 369}, 1 (2006).
  \bibitem{Janssen99} H. K. Janssen{,} K.  Oerding{,} F. van Wijland  and H. Hilhorst, Eur. Phys. J. B {\bf 7}, 139 (1999).
  \bibitem{Hinrichsen07} H. Hinrichsen, J. Stat. Mech., P07006  (2007).
  \bibitem{Janssen97} H. K. Janssen, Phys. Rev. E {\bf 55}, 6253 (1997).
  \bibitem{Kra68} R. Kraichnan, Phys. Fluids {\bf 11}, 945 (1968).
  \bibitem{Ant99} N. V. Antonov, Phys. Rev. E {\bf 60}, 6691 (1999). 
  \bibitem{FGV01} G. Falkovich, K. Gawedzki and M. Vergassola, Rev. Modern Phys. {\bf 73}, 913 (2001).  
  \bibitem{Janssen76} H.~K.~Janssen, Z. Phys. B: Condens. Matter {\bf 23}, 377 (1976).
  \bibitem{deDom76} C.~De~Dominicis, J. Phys. Colloq. France {\bf 37}, C1-247 (1976).
  \bibitem{Janssen79} H.~K.~Janssen, {\it Dynamical Critical Phenomena and Related Topics}, Lect. Notes Phys. {\bf 104}, (Springer, Heidelberg, 1979).
 \bibitem{Antonov06} N. V. Antonov, M. Hnatich, J. Honkonen, J. Phys. A: Math. Gen. {\bf 39}, 7867 (2006).
 \bibitem{AntKap08} N.~V.~Antonov, V.~I.~Iglovikov, A.~S.~Kapustin, J. Phys. A: Math. Theor. {\bf 42}, 135001 (2008).
 \bibitem{Landau} L.~D.~Landau and E.~M.~Lifshitz, {\it Fluid Mechanics} (Pergamon Press, 1959).
  \bibitem{HHL} M.~Hnati\v{c}, J. Honkonen, T. Lu\v{c}ivjansk\'y, Acta Physica Slovaca {\bf 66}, 69 (2016).
  \bibitem{Collins} J.C. Collins, {\it Renormalization}, Cambridge University Press, Cambridge (1984).  
\end{thebibliography}
\end{document}